\documentclass{ws-mpla}
\usepackage[super]{cite}
\usepackage{graphicx}
\usepackage{amsmath}
\usepackage{amssymb}
\begin{document}

\markboth{Oem Trivedi}{Rejuvenating the hope of a swampland consistent inflated multiverse with tachyonic inflation in the high energy RS-II Braneworld
}

\catchline{}{}{}{}{}

\title{ Rejuvenating the hope of a swampland consistent inflated multiverse with tachyonic inflation in the high energy RS-II Braneworld
}

\author{Oem Trivedi\footnote{oem.t@ahduni.edu.in}}

\address{School of Arts and Sciences, Ahmedabad University\\
Ahmedabad 3800009, Gujarat,\\
oem.t@ahduni.edu.in}

\maketitle

\pub{Received (Day Month Year)}{Revised (Day Month Year)}

\begin{abstract}
The swampland conjectures from string theory have had some really interesting implications on cosmology, in particular on inflationary models. Some models of inflation have been shown to be incompatible with these criterion while some have been shown to be severely fine tuned, with most of these problems arising in single field inflationary models in a General relativistic cosmology. Recent works have although optimistically shown that single field models in more general cosmologies can be consistent with these conjectures and hence there is an optimism that not all such models lie in the swampland. However a paradigm of inflation which has been shown to not be perfectly okay with the conjectures is eternal inflation. So in this work, we discuss Tachyonic inflation in the high energy RS-II Braneworld scenario in the context of the swampland conjectures while also considering the possibility of swampland consistent eternal inflation. We show that our concerned regime evades all the prominent swampland issues for single field inflation being virtually unscathed. After this, we show that the main conflicts of eternal inflation with the swampland can easily be resolved in the considered tachyonic scenario and in particular, we also discuss the exciting prospect of a Generalized Uncertainty Principle facilitating the notion of  Swampland consistent eternal inflation. Our work as a whole reignites the possibility that there can be a swampland (and possibly, quantum gravitationally) consistent picture of a "Multiverse".

\keywords{Eternal infation; Swampland conjectures; Tachyonic scalar field}
\end{abstract}

\ccode{PACS Nos.: include PACS Nos.}

\section{Introduction}
The idea of Cosmic Inflation has achieved a tremendous amount of success in describing various properties of the early universe \cite{starobinskii1979spectrum,sato1981first,guth1981inflationary,linde1983chaotic,linde1995quantum}. Numerous predictions of Inflation for the early universe have been repeatedly validated by various satellite experiments, and the most recent data from the Planck experiment follows this trend \cite{aghanim2020planck,akrami2020planck1,aghanim2018planck,akrami2018planck}. Further the observational data supports a huge variety of Inflationary models which are motivated from vastly different backgrounds, from Modified gravity theories to quantum gravitational realizations \cite{martin2014encyclopaedia,martin2014best,wands2008multiple,berera1995warm,ashtekar2010loop,langlois2008perturbations,golovnev2008vector,setare2013warm,kanti2015gauss,dvali1999brane,alexander2013chern}. Inflation has not only been thoroughly studied in regimes where the inflaton is a scalar or a vector field, but also for very non standard scenarios as well where the Inflaton is a complex scalar, a tensor or even a tachyonic field \cite{li2002tachyon,aashish2018inflation,nozari2013some,bilic2019tachyon,gao2018constant,herrera2006tachyon,motaharfar2016warm,mohammadi2020warm,kamali2017tachyon,rezazadeh2017tachyon,scialom1996inflation,Barbosa-Cendejas:2017vgm,el2006phase,bernardini2016entropic,bernardini2017configurational,el2008effective,el2009maxwell,pavvsivc2013localized,das2010warm,rami2008accelerated,herrera2014localization,ahmad2013exact,xiao2019tachyonic,morikawa2022quantum,schwartz2018tachyon,fanchi2022tachyon} . A form of Inflation which has attained widespread interest in the cosmological community in recent decades is Eternal inflation \cite{steinhardt1982natural,vilenkin1983birth,vilenkin2000eternal,guth2000inflation,borde1994eternal,guth2007eternal,linde1986eternal,khoury2019accessibility,jain2019statistics,jain2020eternal,rudelius2019conditions,hawking2018smooth}. Perhaps the most starkling outcome of Inflation continuing eternally is the production of a "Multiverse", as Inflation does not have to stop everywhere at once and can keep on happening in some parts of space while it ceases in some other parts. 
\\
\\
There has been a lot of work dedicated towards a " Theory of Everything " in recent years, and arguably the most well known candidate for such a paradigm is String Theory \cite{green1981supersymmetrical,green1982supersymmetric,green1982supersymmetric1,gubser1998gauge,seiberg1999string,garfinkle1991charged,susskind2003anthropic,polchinski1998string,lust1989lectures}. As String theory presents itself in such a vivid mannerism, one can reasonably expect this theory to have wide ranging implications for cosmology. Consequently, there is a rich and diverse amount of literature which has explored the cosmological implications of the ideas of String theory \cite{mcallister2008string,gasperini1994dilaton,gasperini2003pre,gasperini2007elements,tseytlin1992elements,krori1990some,heckman2019f,heckman2019pixelated}. Amongst the many cosmologically intriguing  features of string theory is the incredibly high amount of possible vacua states it allows, which goes as high as $\mathcal{O} (10^{500})$ and this goes to constitute what is known as the "landscape" of string theory. A natural question which then arises is exactly what class of low energy effective field theories are actually consistent with String theory. In a bid to answer this question, Vafa introduced the term "Swampland" to refer to the class of low energy effective field theories which are inconsistent with the framework of String theory.  \cite{vafa2005string}. Further, in recent years a number of field theoretic UV completion criterion from string theory called the  "swampland conjectures " have been proposed \cite{obied2018sitter,ooguri2016non,garg2019bounds,mcnamara2019cobordism,bedroya2019trans} which classify whether a given regime lies in the swampland or not. As string theory is also seen by many as a viable paradigm of quantum gravity, if a low energy EFT satisfies these criterion then it could also potentially be on amicable terms with a self-consistent theory of quantum gravity. While there have been a considerable number of swampland conjectures (and their modifications) proposed in recent times, the conjectures which have had very telling cosmological implications are :
\\
\\
$ 1 $ : Swampland Distance conjecture (SDC) : This conjecture limits the field space of validity of any effective field theory \cite{ooguri2016non} . This sets a maximum range traversable by the scalar fields in an EFT as \begin{equation}
\frac{\Delta \phi}{m_{p}} \leq d \sim \mathcal{O} (1)
\end{equation} 
where $ m_{p} $ is the reduced Planck's constant, d is some constant of $ \mathcal{O} (1) $ , and $\phi$ is the Scalar Field of the EFT. 
\\
\\
$ 2 $ Swampland De Sitter conjecture (SDSC) : This Conjecture states that it is not possible to create dS Vacua in String Theory \cite{obied2018sitter}. This conjecture is a result of the observation that it has been very hard to generate dS Vacua in String Theory \cite{dasgupta2019sitter,danielsson2018if}( While it has been shown that creating dS Vacua in String Theory is possible in some schemes ,like the KKLT Construction \cite{kachru2003sitter}). The Conjecture sets a lower bound on the gradient of Scalar Potentials in an EFT , \begin{equation}
m_{p}  \frac{| V^{\prime} |}{V} \geq c \sim \mathcal{O} (1)
\end{equation} 
where c is some constant of $ \mathcal{O} (1) $ , and V is the scalar Field Potential. Another refined form of the Swampland De Sitter Conjecture (RSDSC) places constraints on the hessian of the scalar potential (a finding which first appeared in \cite{garg2019bounds} and later in \cite{ooguri2019distance} ) and is given by \begin{equation}
m_{p}   \frac{ V^{\prime \prime}}{V} \leq - c^{\prime} \sim \mathcal{O} (1)
\end{equation} 
\\
\\
These conjectures have pretty interesting implications on cosmology and in particular on single field inflation. It was shown in \cite{kinney2019zoo} that single field inflation in a GR based cosmology is not consistent with these conjectures, considering the data on Inflation \cite{aghanim2020planck,akrami2020planck1}. A lot of work has been done to alleviate this conflict \cite{geng2020potential,scalisi2019swampland,scalisi2020inflation,ashoorioon2019rescuing}. It has ,however, been shown that if the background cosmology for single field inflation is not GR based then this regime of inflation can still be consistent with the swampland criterion \cite{lin2019chaotic,odintsov2020swampland,blumenhagen2017swampland,yi2019gauss}. Particularly, in \cite{trivedi2020swampland} it was shown that (cold) single field inflation can be consistent with these criterion in a large class of non-GR based cosmologies. It's also worth noting that the paradigm of warm inflation \cite{berera1995warm} has also been shown to be quite consistent with the swampland criterion for single models in both GR and non-GR based cosmologies \cite{motaharfar1810warm,motaharfar2019warm,das2019note,das2019warm,das2020runaway,das2020swampland,kamali2020warm}. Another swampland conjecture which has gathered quite an immediate interest with regards to inflationary cosmology is the recently proposed "Trans Planckian Censorship Conjecture"(TCC) \cite{bedroya2019trans}. It was shown in \cite{bedroya2020trans} that single field GR based inflationary models can only be consistent with the TCC if they are severely fine tuned, which is quite ironic considering that inflation was made to solve the fine tuning problem in standard big bang cosmology. A lot of work has further been done to understand the issues of the TCC with single field inflation \cite{brahma2020trans,brandenberger2020strengthening,dhuria2019trans,kamali2020relaxing,bernardo2020trans,mizuno2020universal,lin2019chaotic}. In \cite{brahma2020trans1}, it also shown that the TCC can actually be derived from the distance conjecture (1) considering that latter criterion is true. Single field models in non standard inflationary regimes have been shown to not be severely fine tuned due to the TCC, unlike their counterparts models in a GR based cosmology \cite{berera2019trans,adhikari2020power}. The current literature on inflation and the swampland criterion hence suggests that these conjectures support the notion that early universe expansion happened in a non standard inflationary regime, considering that inflation was the cause of the expansion (and also with the added consideration of only single field models, because multi field models have been shown to be on a very good standing with the conjectures in even the standard paradigm \cite{bravo2020tip,aragam2020multi}). 
\\
\\
While it has been shown that some paradigms of inflation can be rescued from the swampland with different schemes, eternal inflation has been one peculiar inflationary regime which has not been fully consistent with all the swampland conjectures till now. The conflicts of eternal inflation and the swampland were first highlight by Matsui and Takahashi in \cite{matsui2019eternal}, where they showed that one of the core requirements for eternal inflation was in direct conflict with the dS conjecture (2). Dimopoulos then considered eternal inflation on steep potentials with a turning point at $ \dot{\phi} = 0 $ (where $\phi$ is the inflaton field) and found that these models of eternal inflation were also in conflict with the dS conjecture. It was then shown by Kinney \cite{kinney2019eternal} that eternal inflation can be consistent with the Refined dS conjecture (3), and this could prevent eternal inflation from being in the swampland( as it has been shown that consistency with either of (2) or (3), if not both, can prevent a paradigm from plunging into the swampland \cite{garg2019bounds} ). But it was then shown by Brahma and Shandera in \cite{brahma2019stochastic} that eternal inflation can actually not be consistent with the Refined dS conjecture as well. In addition to this, Wang et.al \cite{wang2019eternal} showed that the Gibbons-Hawking entropy bounds for eternal inflation are also not consistent with the swampland conjectures , which is again a very serious issue for eternal inflation. Hence, it does look like eternal inflation has some unavoidable conflicts with the conjectures. However, a key point all of the ventures discussed above is that they considered eternal inflation in a GR based cosmology.The situation does look a tad bit better in more general cosmologies , as Lin et.al showed in \cite{lin2020topological} that certain hilltop models of eternal inflation in a braneworld cosmology can be consistent with the dS conjecture and its refined form. This is encouraging and could point towards the idea that in order for eternal inflation to be consistent with the swampland conjectures, one should look towards regimes which are motivated by String theory (as braneworld cosmology also has its roots in string theory \cite{randall1999large,randall1999alternative,gogberashvili2002hierarchy}). 
\\
\\
A regime of inflation which was inspired from string theory itself is tachyonic inflation \cite{li2002tachyon}, where the inflaton field is considered to be of tachyonic nature. This paradigm of inflation has been shown to be consistent with the swampland criterion in a GR based cosmology for a warm inflationary setup \cite{mohammadi2020warm} under some conditions. Tachyonic inflation has also been studied in a braneworld background cosmology \cite{kamali2016tachyon,kamali2017tachyon,bilic2019tachyon}, so it would be quite interesting to see if tachyonic inflation on the RS-II Brane is consistent with the conjectures, as these scenarios for single field cases have always been on relatively amicable terms with the conjectures as compared to GR based paradigms. Further,  eternal inflation with tachyonic fields in the braneworld can present a great chance of being consistent with the conjectures and the reasons for that are two-fold. First, braneworld inflationary scenarios have proven to be a great fit with the swampland conjectures in a diverse range of regimes and this cosmological setup, as noted above, traces its roots in string theory itself. Secondly, tachyonic models of inflation are also inspired by string theoretic properties and hence, one can be optimistic in exploring the consistency of such a string influenced regime with the conjectures for eternal inflation. Indeed, in our paper we show that this is indeed the case and hence, this paper is structured as follows. In section II, we discuss some of the basic features of tachyonic inflation after which we show that tachyonic inflationary models in the high energy RS-II Braneworld are virtually unscathed by the swampland conjectures and are consistent with all of the criterion individually too. Then, in Section III we show that eternal inflation with tachyonic scalar fields in a braneworld scenario suffers no issues with the swampland conjectures and finally in Section IV, we conclude with some final remarks on our explorations.
\\
\\
\section{Tachyonic Inflation in the RSII Braneworld and the swampland criterion }

We start with the 4-dimensional action of the tachyon field minimally coupled to gravity, which can be written as \cite{li2002tachyon,nozari2013some} \begin{equation}
S = \int \Big[  {m_{p}}^{2} \frac{R}{2} - V(\phi) \sqrt{1 - g^{\mu \nu} \partial_{\mu} \phi \partial_{\nu} \phi } \Big] \sqrt{-g} d^{4} x
\end{equation}
The stress energy tensor in a spatially flat FLRW metric $ ds^{2} = -dt^{2} + a(t)^{2} d \boldsymbol{x}^{2} $ is given by \begin{equation}
{T_{\nu}^{\mu}} = \frac{\partial \mathcal{L}}{\partial (\partial_{\mu} \phi)} \partial_{\nu} \phi - {g_{\nu}^{\mu}} \mathcal{L} = diag (-\rho_{\phi},\tilde{p_{\phi}})
\end{equation} 
where $ \mathcal{L} = \sqrt{-g} \big[{m_{p}}^{2} \frac{R}{2} - V(\phi) \sqrt{1 - g^{\mu \nu} \partial_{\mu} \phi \partial_{\nu} \phi } \big] $ is the Lagrangian density of the Tachyon field , $ V(\phi) $ is the potential,  while $ \rho_{\phi} $ and $ p_{\phi} $ are the energy and pressure densities of the field given by \begin{equation}
\rho_{\phi} = \frac{V(\phi)}{\sqrt{1 - \dot{\phi}^{2}}}
\end{equation}
\begin{equation}
p_{\phi} = - V(\phi) \sqrt{1 - \dot{\phi}^{2}}
\end{equation}
The Friedmann equation for a Randall-Sundrum II Braneworld cosmological scenario is given by [ for details on this, please refer to \cite{ida2000brane,mukohyama2000brane,sahni2002new,brax2004brane}] \begin{equation}
H^{2}  = \frac{\rho}{3 m_{p}^{2}} \left[ 1 + \frac{\rho}{2 \lambda}  \right] + \frac{\Lambda_{4}}{3} + \frac{\omega}{a^{4}}
\end{equation}
where we are working in $ c = \hbar =  1 $ units, with $\rho$ being the energy density, $\lambda$ being the brane tension , $ \Lambda_{4} $ being the 5D cosmological constant and $\omega$ being the so called "dark radiation " term. As we are considering the very early universe, we can ignore the cosmological constant term, whereas the dark radiation term vanishes rapidly due to the $ a^{-4} $ dependence. So, we can write the friedmann equation during inflation as \begin{equation}
H^{2} = \frac{\rho_{\phi}}{3 m_{p}^{2}} \left[ 1 + \frac{\rho_{\phi}}{2 \lambda}  \right] = \frac{1}{3 m_{p}^{2}} \frac{V(\phi)}{\sqrt{1 - \dot{\phi}^{2}}} \left[ 1 + \frac{1}{2 \lambda} \frac{V(\phi)}{\sqrt{1 - \dot{\phi}^{2}}} \right]
\end{equation}
From the action, one can arrive at the field equation of motion as \cite{nozari2013some}, \begin{equation}
\frac{\ddot{\phi}}{1 - \dot{\phi}^{2}} + 3 H \dot{\phi} + \frac{V^{\prime}}{V} = 0
\end{equation}
while one can also use energy conservation to write , \begin{equation}
\dot{\rho_{\phi}} + 3 H (\rho_{\phi} + p_{\phi}) = 0 
\end{equation}
In the slow roll limit of inflation with tachyonic scalar fields, we have $ \dot{\phi} << 1 $ and $ \ddot{\phi} << 3 H \dot{\phi} $ , which allows us to write the Friedmann equation and the field equation of motion as, \begin{equation}
H^{2} = \frac{V}{3 m_{p}^{2}} \Big[1 + \frac{V}{2 \lambda} \Big]
\end{equation}
\begin{equation}
3 H \dot{\phi} + \frac{V^{\prime}}{V} \approxeq 0 
\end{equation}
it's immediately apparent that (12) reduces to the usual Friedmann equation for a GR based cosmology $ H^2 = \frac{V}{3m_{p}^{2}} $ in the low energy limit $ \lambda >> V $. Hence, to better illustrate the effects of the braneworld scenario for inflation, one can instead consider the high energy limit of the picture $ V >> \lambda $, which allows us to now write the Friedmann equation (12) as, \begin{equation}
H^{2} = \frac{V^{2}}{6 \lambda m_{p}^{2}}
\end{equation}
The Number of e-folds can be written using it's usual definition for some initial field value $ \phi_{i} $ to some final value $ \phi_{e} $ as, \begin{equation}
N = \int_{\phi_{i}}^{\phi_{e}} \frac{H}{\dot{\phi}} d \phi
\end{equation}
And using the approximation (13), we can then write the e-fold number as \begin{equation}
N = \int_{\phi_{e}}^{\phi_{i}} \frac{3 H^{2} V}{V^{\prime}} d \phi
\end{equation}
Further, using the Friedmann equation (14), one can write N as \begin{equation}
N = \int_{\phi_{e}}^{\phi_{i}} \frac{V^{3}}{ 2 \lambda m_{p}^{2} V^{\prime}}  d \phi
\end{equation}
A particularly important class of parameters for any inflationary model are the slow roll parameters \cite{martin2014encyclopaedia} and in particular, the $\epsilon$ and $\eta$ parameters. One can define these parameters for tachyonic brane inflation through their usual definitions in the high energy limit as  , \begin{equation}
\epsilon = - \frac{\dot{H}}{H^{2}} = \frac{{ 2  \lambda m_{p}^{2} V^{\prime}}^{2}}{V^{4}} 
\end{equation}
\begin{equation}
\eta =  - \frac{ \ddot{H}}{H \dot{H}} = 2 \sqrt{6 \lambda} \left[ \frac{1}{V^{\prime}} \frac{m_{p} V^{\prime \prime} }{V} - \frac{1}{V} \frac{m_{p} V^{\prime}}{V}   \right]
\end{equation}
Further the power-spectrum of the curvature perturbation $ P_{R} $  that is derived from the correlation of first order scalar field perturbation in the vacuum state can be written \footnote{At this point it is important to clarify about the applicability of this form of the power spectrum in the model we have considered as one might think that the amplitude needs some corrections. Indeed there are corrections to the power spectrum amplitude when one considers non-GR based scenarios butthe form considered here has been shown to be an appropriate approximation in this regard. In the context of the RS-II braneworld scenario, Einstein Equation in 4D would be modified because of the effect of the extra dimension which might change the scalar perturbation part as well at the background level or Friedman equation. There are indeed modifications in the formula of scalar power spectrum because of the modified first Friedman Equation (12) which can be taken into account easily in (20) ( but we have not done so because the premise of our analysis did not require that).  However, the effect of the modification of the perturbation equation is negligible, and with a good approximation, Eq.(20) is quite okay. It has been used previously for describing tachyonic inflation in the RS-II braneworld successfully, see for example \cite{kamali2017tachyon} where the authors used the same form of the power spectrum to construct a viable tachyonic braneworld inflationary scenario.} as  \cite{kamali2017tachyon,hwang2002cosmological} \begin{equation}
P_{R} (k) = \left( \frac{H^{2}}{2 \pi \dot{\phi}} \right)^{2} \frac{1}{V (1 - \dot{\phi}^{2})}
\end{equation}
where in the slow roll limit, one can ignore the contribution of $ \dot{\phi} $. One can then further find out other observationally revelant perturbation parameters like the scalar and tensor spectral index, tensor-to-scalar ratio etc. using their standard definitions \cite{li2002tachyon,nozari2013some,hwang2002cosmological}. We will not illustrate that here as we have built enough groundwork to start seeing why tachyonic brane inflation is consistent with the swampland in the high energy regime.
The prominent issues of conflict between the swampland conjectures and single field inflation which were discussed in length in \cite{kinney2019zoo} can be briefed as follows. One of the issues concerns the dS conjecture and inflationary requirement for the $\epsilon$ parameter, as it was shown that if one considers the dS conjecture seriously than the requirement that $\epsilon << 1 $ is categorically not satisfied. Another issue concerns the fatal constraints that the dS and distance conjecture applied together have for the e-fold number, as they constrain $ N<<1 $ which is quite clearly in violation of  the bare minimum inflationary requirements on the number of e-folds.Finally, the third main issue which was elaborated in \cite{kinney2019zoo} is the inconsistency between constraints set on the $\eta$ parameter from observational data on inflation \cite{akrami2020planck1} and the implications that the refined dS conjecture has on the same parameter.We will now elaborate under what conditions Tachyonic Inflation on the RS-II Braneworld evades all of the issues of single field inflation \footnote{One could probably ask whether tachyonic branewrold models have any realism attached with it and we would like to address that here. Various inflationary regimes in the RS-II Braneworld with a tachyonic scalar have been shown to be observationally viable, with both cold and warm inflationary models being supported in this regard. Some works have been exclusively based on confronting such models with observations and have had positive results, see for example \cite{kamali2017tachyon,kamali2016tachyon,bilic2019tachyon}.} with the swampland conjectures. Firstly, focusing on. The $\epsilon$ parameter (18), can be written as \begin{equation}
\epsilon = \frac{2 \lambda}{V^{2}} \left( \frac{m_{p} V^{\prime}}{V} \right)^{2} 
\end{equation}
Considering the dS conjecture(2) in this scenario, we can write \begin{equation}
\epsilon > \frac{2 \lambda}{V^{2}} c^{2}
\end{equation}
As $ c \sim \mathcal{O} (1) $ , in order for $ \epsilon << 1 $ , one would require \begin{equation}
V >> \sqrt{2 \lambda}
\end{equation}
Hence, it would appear that the dS conjecture implies a constraint on the energy scale of inflation ( as during inflation $\rho = \rho_{\phi} \approxeq V $ ) for tachyonic inflation on the RS-II braneworld. But, it is far from the case as we recall that the formulation built above is in the high energy limit of the braneworld, which characterized by $ V >> \lambda $. So in reality, tachyonic inflation in the high energy RS-II braneworld is not strongly constrained by the dS conjecture and is virtually unscathed. Similarly, the change in the e-fold number during inflation can be written using (17) as, \begin{equation}
\Delta N \approxeq \frac{V^{2}}{2 \lambda} \left[ \frac{\Delta \phi}{m_{p} \left( \frac{m_{p} V^{\prime}}{V}\right)  } \right]
\end{equation}
Now considering both the distance (1) and dS conjecture , we see that term in the square brackets in (24) has to be less than unity. This shows that in order to have sufficient amount of e-folds, one again needs \begin{equation}
V >> \sqrt{2 \lambda}
\end{equation}
and once again as we are in the high energy limit already( $ V >> \lambda $), the requirement (25) does not pose a very strong constraint on the energy scale of inflation in such a regime which might have otherwise ruled out some inflationary models in this scenario. So this makes it clear that there is no issue of a insufficient e-fold number for this inflationary regime. Finally we note that even if one considers both the dS conjecture and the refined dS conjecture (3) to simultaneously hold true and applies them the on $\eta$ parameter (19),  the observational requirement for $ \eta \leq 0 $ is still identically satisfied as both terms in the square bracket in (19) become negative . In conclusion the only "constraint" applied by the swampland conjectures on tachyonic inflation in the high energy RS-II Braneworld  is a small lower limit on the energy scale of inflation, which in the high energy limit is quite intrinsically satisfied. With this we see that this regime of inflation is strictly not in the swampland for effectively all kinds of potentials.
As we have now demonstrated that our concerned inflationary paradigm is quite consistent with the swampland conjectures, we now proceed towards showing how eternal inflation in this regime is also consistent with these criterion.
\\
\\
\section{Swampland consistent Eternal Inflation}

As discussed in section I, recent works have shown that eternal inflation is not on an amicable standing with the swampland conjectures . One of the key requirements in order for eternal inflation to take place in any kind of regime that quantum fluctuations of the inflaton field should dominate over the classical field evolution. The amplitude of quantum fluctuations on scales of order the Hubble length is given by (the following expression is independent of the underlying gravitational theory \cite{nozari2013some}) \begin{equation}
<\delta \phi>_{_{Q}} = \frac{H}{2 \pi}
\end{equation} 
while the classical field variation in Horizon time $ H^{-1} $ is given by \begin{equation}
\delta \phi_{cl} = \frac{\dot{\phi}}{H}
\end{equation}
The requirement that quantum fluctuations dominate the classical field evolution hence translates into \begin{equation}
\frac{<\delta \phi>_{_{Q}}}{\delta \phi_{cl}} = \frac{H^{2}}{2 \pi \dot{\phi}} > 1 
\end{equation}
In the usual case of single scalar field inflation, the criterion (28) transforms to the following requirement on the amplitude of curvature perturbations at horizon crossing \begin{equation}
P_{R} (k) > 1 
\end{equation}
The above requirement can eventually be shown to be inconsistent with the dS conjecture \cite{matsui2019eternal,dimopoulos2018steep}. This was the first conflict highlighted between the swampland criterion and eternal inflation. It was then shown in \cite{kinney2019zoo} that the curvature perturbation amplitude requirement (29) can be amicable with the refined dS conjecture but it was further shown in \cite{brahma2019stochastic} that the perturbation requirements cannot be consistent with the refined dS conjecture as well . Now writing the amplitude of curvature perturbations for tachyonic inflation (20) in the slow roll limit $ \dot{\phi} << 1 $, one gets \begin{equation}
P_{R} (k) \approxeq \left( \frac{H^{2}}{2 \pi \dot{\phi}} \right)^{2} \frac{1}{V}
\end{equation}
From (30), it is immediately apparent that the condition (28) does not set a direct requirement on the amplitude of curvature perturbations for a tachyonic field like it does for a usual scalar field. Further for a large enough energy scale , just like the high energy limit of the RS-II Braneworld considered here, one can easily have $ P_{R} (k) < 1 $ whilst still preserving the condition (28). Hence the conflicts of the swampland conjectures with eternal inflation in the presence of the requirement of $ P_{R} (k) \geq 1  $ \cite{matsui2019eternal,kinney2019eternal,brahma2019stochastic} are not encountered for our concerned tachyonic regime. 
\\
\\
Another very severe issue faced by eternal inflation is the inconsistency its entropy bounds have with the dS conjecture, something which was highlighted in \cite{wang2019eternal} (we are not considering technical entropy based issues regarding stability and configurational problems as they lie beyond the score of our paper, as we are interested in the swampland issues here ). This issue is in particular focused for large field inflationary models and can be understood by considering the Bekenstein-Gibbons-Hawking entropy of the local event horizon of an expanding cosmology \cite{gibbons1977cosmological,gibbons1977action,bekenstein1973black} \begin{equation}
S_{BGH} = \frac{4 \pi m_{p}^{2}}{H^{2}}
\end{equation}
During a period in stochastic inflation when the inflaton moves up the potential due to quantum fluctuations, H increases and thus the entropy of the area bounded by the horizon decreases. For the process above to be consistent with the second law of thermodynamics even after taking into account quantum gravitational effects, the variation of the entropy above is bounded as \cite{bousso2006eternal} \begin{equation}
\delta S > -1 
\end{equation}
Now if one considers a usual scalar field inflationary model based in a General Relativistic cosmology, then bound (31) eventually turns out to be in violation of the dS conjecture and thus leading to serious issue for the entropy bounds for eternal inflation which was discussed in \cite{wang2019eternal}. This issue of entropy bounds, however, will not persist for Tachyonic inflation in a high energy RS-II braneworld which is what we now show. Using the Friedmann equation (14), one can write the variation of the entropy (31) as \begin{equation}
\delta S = - \frac{48 \pi m_{p}^{4} V^{\prime} \lambda}{V^{3}} \delta \phi 
\end{equation}
further, considering the change in entropy during a time in which using $ \delta \phi = \frac{H}{2 \pi} $, we have \begin{equation}
\delta S  = - \frac{24 m_{p}^{2} \lambda}{V} \frac{m_{p} V^{\prime}}{V} \frac{1}{\sqrt{6 \lambda}}
\end{equation}
Now, employing the criterion (32) leads to the following inequality \begin{equation}
\frac{m_{p} V^{\prime}}{V} < \frac{V}{4 \sqrt{6 \lambda} m_{p}^{2}}
\end{equation}
The dS conjecture sets the limit $ \frac{m_{p} V^{\prime}}{V} \geq \mathcal{O} (1) $, hence in order for the above inequality to be consistent with the dS conjecture we need \begin{equation}
V >> 4 \sqrt{6 \lambda} m_{p}^{2} 
\end{equation}
Hence the dS conjecture sets a lower limit on the energy scale of tachyonic inflation in the high energy RS-II braneworld, for it to be on amicable terms with the entropy bounds. We note that the bound to bypass the entropy issue (36) is in the range with lower limit on the inflation energy scale(25) needed to evade the primary issues of the swampland and single field inflation discussed in Section II. So the crux of the matter here is that in order for the eternal inflation entropy bounds to be consistent with the dS conjecture in our concerned tachyonic regime, we need to set a higher lower bound on the energy scale of inflation which is still consistent with the bounds needed to sort out the primary inflationary issues with the swampland. There is, however, another way to go about the entropy bound issue. The event horizon entropy considered here is through the standard Bekenstein-Gibbons-Hawking entropy formulation, while in recent years there has been a lot of work which has explored corrections needed in (31) in order to accommodate the newly emerging physics from string theory and loop quantum gravity (LQG) \cite{kaul2000logarithmic,ghosh2005improved,medved2004conceptual,amelino2004severe,amelino2006black,meissner2004black,adler2001generalized,ali2009discreteness,parikh2000hawking,majumder2011black,majumder2013black,das2008universality}. Several of these approaches have the quantum-corrected entropy-area relation in a general form in terms of (31) as \begin{equation}
S = S_{BGH} + C_{o}  \ln (S_{BGH}) + \sum_{n=1}^{\infty} C_{n} \left(S_{BGH} \right)^{-n}
\end{equation}
where $ C_{n} $ are parameters which are model dependent (like recent works have determined $C_{o} = - \frac{1}{2} $ for LQG  \cite{meissner2004black}). It was also interestingly pointed out in \cite{mead1964possible} that Heisenberg's Uncertainity principle might be affected by quantum gravitational effects. Since then, a significant amount of work has gone into finding a Generalized Uncertainty Principle(GUP) and we can discuss this work in short here. The existence of a minimal length and a maximum momentum accuracy is preferred by various physical observations. Furthermore, assuming modified dispersion relation allows for a wide range of applications in estimating, for example, the inflationary parameters, Lorentz invariance violation, black hole thermodynamics, Saleker-Wigner inequalities, entropic nature of the gravitational laws, Friedmann equations, minimal time measurement and thermodynamics of the high-energy collisions. One of the higher-order GUP approaches gives predictions for the minimal length uncertainty. Another one predicts a maximum momentum and a minimal length uncertainty, simultaneously. One can consider a GUP of the form \begin{equation}
\Delta x \geq \frac{\hbar}{\Delta p} + \frac{\alpha l_{p}^2 \Delta p}{\hbar}
\end{equation} The above GUP is the generalization of the uncertainty principle which about from minimal length considerations and is known as the Kempf-Mangano-Mann (KMM) GUP. The KMM GUP can be deduced from following commutator \begin{equation}
[\hat{x},\hat{p}] = i \hbar(1 \pm \alpha \hat{p}^{2})
\end{equation}  
where $\alpha$ can be either positive or negative. Its's important to stress though that the magnitude of $\alpha$ has been widely studied in the literature and has been seen to be positive mostly. For instance, in String theory it has been assumed to be of order unity \cite{amati1987superstring,amati1989can,maggiore1994quantum}. From a straightforward analysis of the quantum corrections to the Newtonian potential, one can derive $\alpha = \frac{82 \pi}{5} $ \cite{Scardigli:2016pjs} whereas by studying the deformed Unruh temperature in a maximal acceleration framework one would get $\alpha = \frac{8 \pi^2 }{9} $. Similar results have been found in several wide ranging contexts in things like non-commutative Schwarzschild geometry \cite{Kanazawa:2019llj} and even in the Corpuscular description of Black Holes \cite{Buoninfante:2019fwr}. While the KMM GUP is based on minimal length considerations, one can consider a different GUP of the form \begin{equation}
\Delta x \geq \frac{1}{\Delta p (1 - \beta p^2)}
\end{equation}
where $\beta$ is again a parameter which can take on both positive and negative values. This GUP is due to Pedram \cite{pedram2012higher,pedram2012higher2} and has both minimal length and maximal momentum considerations. Again, the magnitude of $\beta$ has been seen to not be of any extreme magnitudes in the literature. looking for the quantum corrections it provides to the usual Bekenstein-Gibbons-Hawking formulation of local event horizon entropy. This eventually led to the following form of the quantum corrected entropy due to a GUP, \cite{majumder2011black,amelino2006black,adler2001generalized,majumder2013black} 
\begin{equation}
S  = S_{BGH} + \frac{\sqrt{\pi } \alpha_{o}}{4} \sqrt{S_{BGH}} - \frac{\pi \alpha_{o}^{2}}{64}  \ln (S_{BGH}) + \mathcal{O} (\frac{1}{m_{p}^{3}})
\end{equation}
where $ \alpha_{o} $ is dimensionless constant which is found in the deformed commutation relations and can be usually considered to be positive in nature \cite{das2008universality}. The leading order contribution to the GUP corrected entropy formalism is due to the $ \sqrt{S_{BGH}} $ term, which is an extra term to the already existing logarithmic correction to entropy derived from the quantum gravity effects. Considering this form of entropy relation as the one for entropy of a local event horizon for an expanding cosmology has provided pretty interesting insights (see for example, \cite{bandyopadhyay2018thermodynamic}). Hence, we now consider the form (38) of the entropy for the bound (32) and as we would like to entertain in particular the implications of the very interesting GUP effects , we only consider the leading order GUP correction to the entropy which provides us with \begin{equation}
S_{GUP} \approxeq S_{BGH} + \frac{\sqrt{\pi } \alpha_{o}}{4} \sqrt{S_{BGH}} 
\end{equation}
which can be written in terms of the Hubble Parameter as \begin{equation}
S = \frac{m_{p} \pi}{H} \bigg[ \frac{4 m_{p}}{H} + \frac{\alpha_{o}}{2}    \bigg]
\end{equation}
Using the Friedmann equation (14), we can now write the entropy in terms of the potential \begin{equation}
S = \frac{m_{p}^{2} \pi \sqrt{6 \lambda}}{V} \Big[ \frac{4 m_{p}^{2} \sqrt{6 \lambda}}{V} + \frac{\alpha_{o}}{2} \Big]
\end{equation}
This finally allows us to express the bound (32) in terms of the GUP corrected entropy as 
\begin{equation}
\frac{m_{p} V^{\prime}}{V} \frac{1}{2 \pi \sqrt{6 \lambda}} \Bigg[ \frac{\pi \sqrt{6 \lambda}}{V} \left( \frac{4 m_{p}^{2} \sqrt{6 \lambda}}{V} + \frac{\alpha_{o}}{2}  \right) \\ +  \frac{4 m_{p} \sqrt{6 \lambda}}{V} \left( \frac{m_{p}^{2} \pi \sqrt{6 \lambda }}{V} \right)  \Bigg] < 1
\end{equation}
Rewriting the inequality above, we have \begin{equation}
\frac{m_{p} V^{\prime}}{V} < \frac{2 \pi \sqrt{6 \lambda}}{ \frac{48 \pi m_{p}^{2} \lambda}{V^{2}} + \frac{\pi \sqrt{6 \lambda \alpha_{o}}}{2 V}}
\end{equation}
The dS conjecture constraints $ \frac{m_{p} V^{\prime}}{V} \geq c \sim \mathcal{O} (1) $, hence in order to maintain consistency with it the following inequality will have to hold 
\begin{equation}
2 \pi \sqrt{6 \lambda} > \frac{48 \pi m_{p}^{2} \lambda}{V^{2}} + \frac{\pi \sqrt{6 \lambda \alpha_{o}}}{2 V} \\ \implies 2 V^{2} - \frac{V \alpha_{o}}{2} - 8 m_{p}^{2} \sqrt{6 \lambda} > 0 
\end{equation}
The last inequality is a bit non-trivial but can be calculated using computing systems like Mathematica. This inequality sets a lower bounds on $\lambda$ given by  \begin{equation}
\lambda > \frac{\alpha_{o}^{4}}{393216 m_{p}^{4}}
\end{equation} 
And also a lower bound on the inflation energy scale, which keeping in mind the brane tension bound can be written approximately as
\begin{equation}
V >> \frac{\alpha_{o}}{8} 
\end{equation}
Hence the dS conjecture can be consistent with the GUP corrected eternal inflation entropy bounds given that the brane tension and the inflation energy scale abide by the above mentioned lower limits. The most interesting outcome of considering the GUP corrected entropy bounds is in the nature of limits implied in (45-46). If these bounds are satisfied (these are in line with the basic requirements for the energy scale for consistency with the conjectures), then the famous issues which contribute to the failure of a swampland consistent eternal inflation picture are taken care of. Hence the direct relation between the inflaton energy scale and the constant $ \alpha_{o} $ which is found from the deformed commutation relations for a GUP is incredibly exciting. In this scenario, the existence of a Generalized Uncertainty Principle facilitates the creation of a swampland consistent eternal inflationary scenario and consequently, a swampland consistent multiverse. To make this peculiar relation even more clear, we would like to demonstrate that the consideration of a GUP corrected entropy formalism would also allow us to evade the dS conjecture-entropy bound conflict in a simple scalar field model in a GR based scenario. For GR based single field models , the Friedmann equation during the expansion phase is \begin{equation}
H^{2} = \frac{V}{3 m_{p}^{2}}
\end{equation}
The GUP corrected entropy (39) for this regime becomes \begin{equation}
S = m_{p}^{2} \pi \sqrt{\frac{3}{V}}\left( 4 m_{p}^{2} \sqrt{\frac{3}{V}} + \frac{\alpha_{o}}{2}  \right)
\end{equation}
And hence, the bound (32) in this case takes the form \begin{equation}
\frac{m_{p} V^{\prime}}{V} < \frac{8V}{m_{p} H \left[ 48 m_{p}^{2} + \alpha_{o} \sqrt{3V}   \right]} 
\end{equation}
It is worth mentioning the difference the GUP corrected has already made in eradicating the entropy bound issue for this inflationary regime. The above inequality clearly shows that it quite possible to satisfy the constraints due to the dS conjecture on the left hand side of the inequality, given that there is some minimum limit on the energy scale of inflation to make the right hand side of the inequality sufficiently large. Doing this would mean that the numerator on the right hand side far dominates over the denominator which using the Friedmann equation (47), can be written as the following inequality \begin{equation}
V^{1/2} > \frac{12 m_{p}^{2}}{2 - \frac{\alpha_{o} \sqrt{3}}{4}}
\end{equation}
Hence one can eradicate the issue between the dS conjecture and entropy bounds by considering a GUP corrected entropy formalism by constraining the energy scale during inflation in accordance with the above bound for $ \alpha_{o} < \frac{8}{\sqrt{3}} $. Once again, we notice how crucial a role the Generalized Uncertainty Principle based corrections make in allowing us to evade the entropy bound issues even in a usual scalar field model based in a GR based cosmology. This in no way shows that eternal inflation in such a regime is consistent with the swampland, however, as other issues regarding the conjectures still hold true \cite{matsui2019eternal,dimopoulos2018steep,kinney2019eternal,brahma2019stochastic}. But this does illustrate that GUP based quantum corrected entropy does facilitate positively in removing the entropy bound-dS conjecture conflict in multiple regimes of inflation. Thus we can draw a very exciting conclusion from this analysis that considering a generalized uncertainty principle makes it more easier for one to have a swampland consistent (and considering the premise of the conjectures,  a quantum gravitationally consistent) picture of a multiverse generated by eternal inflation.

\section{Concluding remarks and discussion}
In conclusion, in this work we have discussed tachyonic inflation in the high energy RS-II Braneworld scenario in the light of the swampland conjectures and considered the possibility of swampland consistent eternal inflation in this regime. We started off by describing why tachyonic inflation in this particular scenario could possibly be one of the most suited arenas for the swampland conjectures and hence, might allow for the possibility of evading the swampland issues eternal inflation too. Then after describing some basics of Tachyonic inflation in this regime, we showed that this inflationary model is virtually unscathed by the swampland conjectures and can quite intrinsically evade all the prominent swampland issues which have been discussed for regular single field inflation, as they only imply a lower bound on the inflation energy scale which can be easily satisfied in the considered high energy limit. We then cast our glance towards eternal inflation in this scenario, and discussed the two main swampland issues in this context where we firstly addressed the curvature amplitude issue. We showed that the basic eternal inflation requirement of quantum fluctuations dominating over classical field evolution does not set unavoidable constraints on tachyonic models of any kind and in particular, inflation in high energy RS-II Braneworld. We hence showed that the issues of inconsistency of the dS and refined dS conjectures with the perturbation requirement do not hold good for our tachyonic regime. We then discussed the conflict of entropy bounds with the dS conjecture and showed that our concerned inflationary regime does not have any issues with the entropy bound even when the concerned entropy formalism is the standard Bekenstein-Gibbons-Hawking one, as the lower limit on the energy scale implied for consistency with the bound is well within the range needed for consistency with basic inflationary swampland issues. Then made the case as to why it would be better suited if we consider quantum corrections to the BGH entropy , and in particular we considered a Generalized Uncertainty Principle corrected entropy formalism. In order to focus on this particular correction to the entropy, we only considered the leading order GUP corrections to the entropy and showed that a GUP formalism of the entropy allows one to evade the entropy bound issues in both tachyonic inflation in our concerned scenario and even in a usual single field model in a GR based cosmology. This allowed us to make a startling observation that a generalized uncertainty principle actually facilitates one to have a swampland (and hence possibly, a quantum gravitationally) consistent picture of eternal inflation and a Multiverse. An important thing to note is that we have not carried out a detailed analysis of eternal inflation in any kind of a tachyonic regime as we have just focused on showing how the prominent swampland-eternal inflation issues can be resolved in a more quantum gravitationally / String theory motivated setting. Indeed, we have not embarked on exploring the solutions of the Fokker-Planck / Langevin equations for quantum fluctuations of the inflaton field for different potentials, which is more of a standard procedure in the subject. Hence, there might be more model dependent requirements for eternal inflation to take place in a variety for different potentials that we have not taken into account here ( like the model dependent analysis which was done in \cite{brahma2019stochastic}) as our explorations have been of a more general nature focused on the prevalent swampland issues currently.

\section*{Acknowledgements}

The author would like to thank Suddhasatwa Brahma for his insightful comments which led to the initiation of this work. The author would also like to thank Sunny Vagnozzi  for his very helpful advice during the preparation of this manuscript. I would also like to thank the reviewers of the manuscript for their insightful comments on the paper, which has increased the depth of the work by multi folds.

\bibliography{SHREERAMAJI.bib}

\begin{thebibliography}{100}

\bibitem{starobinskii1979spectrum}
AA~Starobinskii.
\newblock Spectrum of relict gravitational radiation and the early state of the
  universe.
\newblock {\em JETP Letters}, 30(11):682--685, 1979.

\bibitem{sato1981first}
Katsuhiko Sato.
\newblock First-order phase transition of a vacuum and the expansion of the
  universe.
\newblock {\em Monthly Notices of the Royal Astronomical Society},
  195(3):467--479, 1981.

\bibitem{guth1981inflationary}
Alan~H Guth.
\newblock Inflationary universe: A possible solution to the horizon and
  flatness problems.
\newblock {\em Physical Review D}, 23(2):347, 1981.

\bibitem{linde1983chaotic}
Andrei~D Linde.
\newblock Chaotic inflation, 1983.

\bibitem{linde1995quantum}
Andrei Linde.
\newblock Quantum cosmology and the structure of inflationary universe.
\newblock {\em arXiv preprint gr-qc/9508019}, 1995.

\bibitem{aghanim2020planck}
Nabila Aghanim, Y~Akrami, F~Arroja, M~Ashdown, J~Aumont, C~Baccigalupi,
  M~Ballardini, AJ~Banday, RB~Barreiro, N~Bartolo, et~al.
\newblock Planck 2018 results.
\newblock {\em Astronomy and Astrophysics-A\&A}, 641:A1, 2020.

\bibitem{akrami2020planck1}
Yashar Akrami, Frederico Arroja, M~Ashdown, J~Aumont, C~Baccigalupi,
  M~Ballardini, AJ~Banday, RB~Barreiro, N~Bartolo, S~Basak, et~al.
\newblock Planck 2018 results-x. constraints on inflation.
\newblock {\em Astronomy \& Astrophysics}, 641:A10, 2020.

\bibitem{aghanim2018planck}
N~Aghanim, Yashar Akrami, M~Ashdown, J~Aumont, C~Baccigalupi, M~Ballardini,
  AJ~Banday, RB~Barreiro, N~Bartolo, S~Basak, et~al.
\newblock Planck 2018 results. vi. cosmological parameters.
\newblock {\em arXiv preprint arXiv:1807.06209}, 2018.

\bibitem{akrami2018planck}
Y~Akrami, F~Arroja, M~Ashdown, J~Aumont, C~Baccigalupi, M~Ballardini,
  AJ~Banday, RB~Barreiro, N~Bartolo, S~Basak, et~al.
\newblock Planck 2018 results. i. overview and the cosmological legacy of
  planck.
\newblock {\em arXiv preprint arXiv:1807.06205}, 2018.

\bibitem{martin2014encyclopaedia}
Jerome Martin, Christophe Ringeval, and Vincent Vennin.
\newblock Encyclop{\ae}dia inflationaris.
\newblock {\em Physics of the Dark Universe}, 5:75--235, 2014.

\bibitem{martin2014best}
J{\'e}r{\^o}me Martin, Christophe Ringeval, Roberto Trotta, and Vincent Vennin.
\newblock The best inflationary models after planck.
\newblock {\em Journal of Cosmology and Astroparticle Physics}, 2014(03):039,
  2014.

\bibitem{wands2008multiple}
David Wands.
\newblock Multiple field inflation.
\newblock In {\em Inflationary cosmology}, pages 275--304. Springer, 2008.

\bibitem{berera1995warm}
Arjun Berera.
\newblock Warm inflation.
\newblock {\em Physical Review Letters}, 75(18):3218, 1995.

\bibitem{ashtekar2010loop}
Abhay Ashtekar and David Sloan.
\newblock Loop quantum cosmology and slow roll inflation.
\newblock {\em Physics Letters B}, 694(2):108--112, 2010.

\bibitem{langlois2008perturbations}
David Langlois and Sebastien Renaux-Petel.
\newblock Perturbations in generalized multi-field inflation.
\newblock {\em Journal of Cosmology and Astroparticle Physics}, 2008(04):017,
  2008.

\bibitem{golovnev2008vector}
Alexey Golovnev, Viatcheslav Mukhanov, and Vitaly Vanchurin.
\newblock Vector inflation.
\newblock {\em Journal of Cosmology and Astroparticle Physics}, 2008(06):009,
  2008.

\bibitem{setare2013warm}
MR~Setare and V~Kamali.
\newblock Warm vector inflation.
\newblock {\em Physics Letters B}, 726(1-3):56--65, 2013.

\bibitem{kanti2015gauss}
Panagiota Kanti, Radouane Gannouji, and Naresh Dadhich.
\newblock Gauss-bonnet inflation.
\newblock {\em Physical Review D}, 92(4):041302, 2015.

\bibitem{dvali1999brane}
Gia Dvali and S-H~Henry Tye.
\newblock Brane inflation.
\newblock {\em Physics Letters B}, 450(1-3):72--82, 1999.

\bibitem{alexander2013chern}
Stephon Alexander, Antonino Marciano, and David Spergel.
\newblock Chern-simons inflation and baryogenesis.
\newblock {\em Journal of Cosmology and Astroparticle Physics}, 2013(04):046,
  2013.

\bibitem{li2002tachyon}
Xin-zhou Li, Dao-jun Liu, and Jian-gang Hao.
\newblock On the tachyon inflation.
\newblock {\em arXiv preprint hep-th/0207146}, 2002.

\bibitem{aashish2018inflation}
Sandeep Aashish, Abhilash Padhy, Sukanta Panda, and Arun Rana.
\newblock Inflation with an antisymmetric tensor field.
\newblock {\em The European Physical Journal C}, 78(11):887, 2018.

\bibitem{nozari2013some}
Kourosh Nozari and Narges Rashidi.
\newblock Some aspects of tachyon field cosmology.
\newblock {\em Physical Review D}, 88(2):023519, 2013.

\bibitem{bilic2019tachyon}
Neven Bili{\'c}, Dragoljub~D Dimitrijevi{\'c}, Goran~S Djordjevic, Milan
  Milo{\v{s}}evi{\'c}, and Marko Stojanovi{\'c}.
\newblock Tachyon inflation in the holographic braneworld.
\newblock {\em Journal of Cosmology and Astroparticle Physics}, 2019(08):034,
  2019.

\bibitem{gao2018constant}
Qing Gao, Yungui Gong, and Qin Fei.
\newblock Constant-roll tachyon inflation and observational constraints.
\newblock {\em Journal of Cosmology and Astroparticle Physics}, 2018(05):005,
  2018.

\bibitem{herrera2006tachyon}
Ram{\'o}n Herrera, Sergio del Campo, and Cuauhtemoc Campuzano.
\newblock Tachyon warm inflationary universe models.
\newblock {\em Journal of Cosmology and Astroparticle Physics}, 2006(10):009,
  2006.

\bibitem{motaharfar2016warm}
Meysam Motaharfar and Hamid~Reza Sepangi.
\newblock Warm-tachyon gauss--bonnet inflation in the light of planck 2015
  data.
\newblock {\em The European Physical Journal C}, 76(11):646, 2016.

\bibitem{mohammadi2020warm}
Abolhassan Mohammadi, Tayeb Golanbari, Haidar Sheikhahmadi, Kosar Sayar, Lila
  Akhtari, MA~Rasheed, and Khaled Saaidi.
\newblock Warm tachyon inflation and swampland criteria.
\newblock {\em Chinese Physics C}, 44(9):095101, 2020.

\bibitem{kamali2017tachyon}
Vahid Kamali and Elahe~Navaee Nik.
\newblock Tachyon logamediate inflation on the brane.
\newblock {\em The European Physical Journal C}, 77(7):449, 2017.

\bibitem{rezazadeh2017tachyon}
K~Rezazadeh, K~Karami, and S~Hashemi.
\newblock Tachyon inflation with steep potentials.
\newblock {\em Physical Review D}, 95(10):103506, 2017.

\bibitem{scialom1996inflation}
David Scialom.
\newblock Inflation with a complex scalar field.
\newblock {\em arXiv preprint gr-qc/9609020}, 1996.

\bibitem{Barbosa-Cendejas:2017vgm}
Nandinii Barbosa-Cendejas, Roberto Cartas-Fuentevilla, Alfredo Herrera-Aguilar,
  Refugio~Rigel Mora-Luna, and Rold\~ao da~Rocha.
\newblock {A de Sitter tachyonic braneworld revisited}.
\newblock {\em JCAP}, 01:005, 2018.

\bibitem{el2006phase}
Ahmad~Rami El-Nablusi.
\newblock Phase transitions in the early universe with negatively induced
  supergravity cosmological constant.
\newblock {\em Chinese Physics Letters}, 23(5):1124, 2006.

\bibitem{bernardini2016entropic}
Alex~E Bernardini and Rold{\~a}o da~Rocha.
\newblock Entropic information of dynamical ads/qcd holographic models.
\newblock {\em Physics Letters B}, 762:107--115, 2016.

\bibitem{bernardini2017configurational}
Alex~E Bernardini, Nelson~RF Braga, and Rold{\~a}o da~Rocha.
\newblock Configurational entropy of glueball states.
\newblock {\em Physics Letters B}, 765:81--85, 2017.

\bibitem{el2008effective}
Ahmad~Rami El-Nabulsi.
\newblock Effective 3-brane brans--dicke cosmology.
\newblock {\em Modern Physics Letters A}, 23(06):401--415, 2008.

\bibitem{el2009maxwell}
RA~El-Nabulsi.
\newblock Maxwell brane cosmology with higher-order string curvature
  corrections, a nonminimally coupled scalar field, dark matter--dark energy
  interaction and a varying speed of light.
\newblock {\em International Journal of Modern Physics D}, 18(02):289--318,
  2009.

\bibitem{pavvsivc2013localized}
Matej Pav{\v{s}}i{\v{c}}.
\newblock Localized propagating tachyons in extended relativity theories.
\newblock {\em Advances in Applied Clifford Algebras}, 23(2):469--495, 2013.

\bibitem{das2010warm}
Pranita Das and Atri Deshamukhya.
\newblock Warm tachyonic inflation in strong dissipative regime: A study with
  exp [-t2] potential.
\newblock {\em Indian Journal of Physics}, 84(6):617--622, 2010.

\bibitem{rami2008accelerated}
El-Nabulsi~Ahmad Rami.
\newblock Accelerated d-dimensional compactified universe in
  gauss--bonnet--dilatonic scalar gravity from d-brane/m-theory.
\newblock {\em Chinese Physics Letters}, 25(8):2785, 2008.

\bibitem{herrera2014localization}
Alfredo Herrera-Aguilar, Alma~D Rojas, and El{\'\i} Santos.
\newblock Localization of gauge fields in a tachyonic de sitter thick
  braneworld.
\newblock {\em The European Physical Journal C}, 74(4):1--6, 2014.

\bibitem{ahmad2013exact}
Rami Ahmad El-Nabulsi.
\newblock Exact solution of a tachyon oscillating cosmology with a supergravity
  tracking potential.
\newblock {\em The European Physical Journal Plus}, 128(5):1--11, 2013.

\bibitem{xiao2019tachyonic}
Kui Xiao.
\newblock Tachyonic inflation in loop quantum cosmology.
\newblock {\em The European Physical Journal C}, 79(12):1--8, 2019.

\bibitem{morikawa2022quantum}
Masahiro Morikawa.
\newblock Quantum fluctuations in vacuum energy: Cosmic inflation as a
  dynamical phase transition.
\newblock {\em Universe}, 8(6):295, 2022.

\bibitem{schwartz2018tachyon}
Charles Schwartz.
\newblock Tachyon dynamics—for neutrinos?
\newblock {\em International Journal of Modern Physics A}, 33(10):1850056,
  2018.

\bibitem{fanchi2022tachyon}
John~R Fanchi.
\newblock Tachyon behavior due to mass-state transitions at scattering
  vertices.
\newblock {\em Physics}, 4(1):217--228, 2022.

\bibitem{steinhardt1982natural}
Paul~Joseph Steinhardt.
\newblock Natural inflation.
\newblock {\em The very early universe}, page 251, 1982.

\bibitem{vilenkin1983birth}
Alexander Vilenkin.
\newblock Birth of inflationary universes.
\newblock {\em Physical Review D}, 27(12):2848, 1983.

\bibitem{vilenkin2000eternal}
Alexander Vilenkin.
\newblock Eternal inflation and the present universe.
\newblock {\em Nuclear Physics B-Proceedings Supplements}, 88(1-3):67--74,
  2000.

\bibitem{guth2000inflation}
Alan~H Guth.
\newblock Inflation and eternal inflation.
\newblock {\em Physics Reports}, 333:555--574, 2000.

\bibitem{borde1994eternal}
Arvind Borde and Alexander Vilenkin.
\newblock Eternal inflation and the initial singularity.
\newblock {\em Physical Review Letters}, 72(21):3305, 1994.

\bibitem{guth2007eternal}
Alan~H Guth.
\newblock Eternal inflation and its implications.
\newblock {\em Journal of Physics A: Mathematical and Theoretical},
  40(25):6811, 2007.

\bibitem{linde1986eternal}
Andrei~D Linde.
\newblock Eternal chaotic inflation.
\newblock {\em Modern Physics Letters A}, 1(02):81--85, 1986.

\bibitem{khoury2019accessibility}
Justin Khoury.
\newblock Accessibility measure for eternal inflation: Dynamical criticality
  and higgs metastability.
\newblock {\em arXiv preprint arXiv:1912.06706}, 2019.

\bibitem{jain2019statistics}
Mudit Jain and Mark~P Hertzberg.
\newblock Statistics of inflating regions in eternal inflation.
\newblock {\em Physical Review D}, 100(2):023513, 2019.

\bibitem{jain2020eternal}
Mudit Jain and Mark~P Hertzberg.
\newblock Eternal inflation and reheating in the presence of the standard model
  higgs field.
\newblock {\em Physical Review D}, 101(10):103506, 2020.

\bibitem{rudelius2019conditions}
Tom Rudelius.
\newblock Conditions for (no) eternal inflation.
\newblock {\em Journal of Cosmology and Astroparticle Physics}, 2019(08):009,
  2019.

\bibitem{hawking2018smooth}
Stephen~W Hawking and Thomas Hertog.
\newblock A smooth exit from eternal inflation?
\newblock {\em Journal of High Energy Physics}, 2018(4):147, 2018.

\bibitem{green1981supersymmetrical}
Michael~B Green and John~H Schwarz.
\newblock Supersymmetrical dual string theory.
\newblock {\em Nuclear Physics B}, 181(3):502--530, 1981.

\bibitem{green1982supersymmetric}
Michael~B Green and John~H Schwarz.
\newblock Supersymmetric dual string theory:(iii). loops and renormalization.
\newblock {\em Nuclear Physics B}, 198(3):441--460, 1982.

\bibitem{green1982supersymmetric1}
Michael~B Green and John~H Schwarz.
\newblock Supersymmetric dual string theory:(ii). vertices and trees.
\newblock {\em Nuclear Physics B}, 198(2):252--268, 1982.

\bibitem{gubser1998gauge}
Steven~S Gubser, Igor~R Klebanov, and Alexander~M Polyakov.
\newblock Gauge theory correlators from non-critical string theory.
\newblock {\em Physics Letters B}, 428(1-2):105--114, 1998.

\bibitem{seiberg1999string}
Nathan Seiberg and Edward Witten.
\newblock String theory and noncommutative geometry.
\newblock {\em Journal of High Energy Physics}, 1999(09):032, 1999.

\bibitem{garfinkle1991charged}
David Garfinkle, Gary~T Horowitz, and Andrew Strominger.
\newblock Charged black holes in string theory.
\newblock {\em Physical Review D}, 43(10):3140, 1991.

\bibitem{susskind2003anthropic}
Leonard Susskind.
\newblock The anthropic landscape of string theory.
\newblock {\em Universe or multiverse}, pages 247--266, 2003.

\bibitem{polchinski1998string}
Joseph Polchinski.
\newblock {\em String theory: Volume 2, superstring theory and beyond}.
\newblock Cambridge university press, 1998.

\bibitem{lust1989lectures}
Dieter L{\"u}st and Stefan Theisen.
\newblock {\em Lectures on string theory}, volume 346.
\newblock Springer, 1989.

\bibitem{mcallister2008string}
Liam McAllister and Eva Silverstein.
\newblock String cosmology: a review.
\newblock {\em General Relativity and Gravitation}, 40(2-3):565--605, 2008.

\bibitem{gasperini1994dilaton}
M~Gasperini and Gabriele Veneziano.
\newblock Dilaton production in string cosmology.
\newblock {\em Physical Review D}, 50(4):2519, 1994.

\bibitem{gasperini2003pre}
Maurizio Gasperini and Gabriele Veneziano.
\newblock The pre-big bang scenario in string cosmology.
\newblock {\em Physics Reports}, 373(1-2):1--212, 2003.

\bibitem{gasperini2007elements}
Maurizio Gasperini.
\newblock {\em Elements of string cosmology}, volume~36.
\newblock Cambridge University Press Cambridge, 2007.

\bibitem{tseytlin1992elements}
Arkady~A Tseytlin and C~Vafa.
\newblock Elements of string cosmology.
\newblock {\em Nuclear Physics B}, 372(1-2):443--466, 1992.

\bibitem{krori1990some}
KD~Krori, T~Chaudhury, Chandra~Rekha Mahanta, and Ajanta Mazumdar.
\newblock Some exact solutions in string cosmology.
\newblock {\em General Relativity and Gravitation}, 22(2):123--130, 1990.

\bibitem{heckman2019f}
Jonathan~J Heckman, Craig Lawrie, Ling Lin, and Gianluca Zoccarato.
\newblock F-theory and dark energy.
\newblock {\em Fortschritte der Physik}, 67(10):1900057, 2019.

\bibitem{heckman2019pixelated}
Jonathan~J Heckman, Craig Lawrie, Ling Lin, Jeremy Sakstein, and Gianluca
  Zoccarato.
\newblock Pixelated dark energy.
\newblock {\em Fortschritte der Physik}, 67(11):1900071, 2019.

\bibitem{vafa2005string}
Cumrun Vafa.
\newblock The string landscape and the swampland.
\newblock {\em arXiv preprint hep-th/0509212}, 2005.

\bibitem{obied2018sitter}
Georges Obied, Hirosi Ooguri, Lev Spodyneiko, and Cumrun Vafa.
\newblock de sitter space and the swampland.
\newblock {\em arXiv preprint arXiv:1806.08362}, 2018.

\bibitem{ooguri2016non}
Hirosi Ooguri and Cumrun Vafa.
\newblock Non-supersymmetric ads and the swampland.
\newblock {\em arXiv preprint arXiv:1610.01533}, 2016.

\bibitem{garg2019bounds}
Sumit~K Garg and Chethan Krishnan.
\newblock Bounds on slow roll and the de sitter swampland.
\newblock {\em Journal of High Energy Physics}, 2019(11):75, 2019.

\bibitem{mcnamara2019cobordism}
Jacob McNamara and Cumrun Vafa.
\newblock Cobordism classes and the swampland.
\newblock {\em arXiv preprint arXiv:1909.10355}, 2019.

\bibitem{bedroya2019trans}
Alek Bedroya and Cumrun Vafa.
\newblock Trans-planckian censorship and the swampland.
\newblock {\em arXiv preprint arXiv:1909.11063}, 2019.

\bibitem{dasgupta2019sitter}
Keshav Dasgupta, Maxim Emelin, Mir~Mehdi Faruk, and Radu Tatar.
\newblock De sitter vacua in the string landscape.
\newblock {\em arXiv preprint arXiv:1908.05288}, 2019.

\bibitem{danielsson2018if}
Ulf~H Danielsson and Thomas~Van Riet.
\newblock What if string theory has no de sitter vacua?
\newblock {\em International Journal of Modern Physics D}, 27(12):1830007,
  2018.

\bibitem{kachru2003sitter}
Shamit Kachru, Renata Kallosh, Andrei Linde, and Sandip~P Trivedi.
\newblock De sitter vacua in string theory.
\newblock {\em Physical Review D}, 68(4):046005, 2003.

\bibitem{ooguri2019distance}
Hirosi Ooguri, Eran Palti, Gary Shiu, and Cumrun Vafa.
\newblock Distance and de sitter conjectures on the swampland.
\newblock {\em Physics Letters B}, 788:180--184, 2019.

\bibitem{kinney2019zoo}
William~H Kinney, Sunny Vagnozzi, and Luca Visinelli.
\newblock The zoo plot meets the swampland: mutual (in) consistency of
  single-field inflation, string conjectures, and cosmological data.
\newblock {\em Classical and quantum gravity}, 36(11):117001, 2019.

\bibitem{geng2020potential}
Hao Geng.
\newblock A potential mechanism for inflation from swampland conjectures.
\newblock {\em Physics Letters B}, page 135430, 2020.

\bibitem{scalisi2019swampland}
Marco Scalisi and Irene Valenzuela.
\newblock Swampland distance conjecture, inflation and $\alpha$-attractors.
\newblock {\em Journal of High Energy Physics}, 2019(8):160, 2019.

\bibitem{scalisi2020inflation}
Marco Scalisi.
\newblock Inflation, higher spins and the swampland.
\newblock {\em Physics Letters B}, 808:135683, 2020.

\bibitem{ashoorioon2019rescuing}
Amjad Ashoorioon.
\newblock Rescuing single field inflation from the swampland.
\newblock {\em Physics Letters B}, 790:568--573, 2019.

\bibitem{lin2019chaotic}
Chia-Min Lin, Kin-Wang Ng, and Kingman Cheung.
\newblock Chaotic inflation on the brane and the swampland criteria.
\newblock {\em Physical Review D}, 100(2):023545, 2019.

\bibitem{odintsov2020swampland}
SD~Odintsov and VK~Oikonomou.
\newblock Swampland implications of gw170817-compatible einstein-gauss-bonnet
  gravity.
\newblock {\em Physics Letters B}, page 135437, 2020.

\bibitem{blumenhagen2017swampland}
Ralph Blumenhagen, Irene Valenzuela, and Florian Wolf.
\newblock The swampland conjecture and f-term axion monodromy inflation.
\newblock {\em Journal of High Energy Physics}, 2017(7):145, 2017.

\bibitem{yi2019gauss}
Zhu Yi and Yungui Gong.
\newblock Gauss--bonnet inflation and the string swampland.
\newblock {\em Universe}, 5(9):200, 2019.

\bibitem{trivedi2020swampland}
Oem Trivedi.
\newblock Swampland conjectures and single field inflation in modified
  cosmological scenarios.
\newblock {\em arXiv preprint arXiv:2008.05474}, 2020.

\bibitem{motaharfar1810warm}
Meysam Motaharfar, Vahid Kamali, and Rudnei~O Ramos.
\newblock Warm way out of the swampland.
\newblock {\em arXiv preprint arXiv:1810.02816}.

\bibitem{motaharfar2019warm}
Meysam Motaharfar, Vahid Kamali, and Rudnei~O Ramos.
\newblock Warm inflation as a way out of the swampland.
\newblock {\em Physical Review D}, 99(6):063513, 2019.

\bibitem{das2019note}
Suratna Das.
\newblock Note on single-field inflation and the swampland criteria.
\newblock {\em Physical Review D}, 99(8):083510, 2019.

\bibitem{das2019warm}
Suratna Das.
\newblock Warm inflation in the light of swampland criteria.
\newblock {\em Physical Review D}, 99(6):063514, 2019.

\bibitem{das2020runaway}
Suratna Das and Rudnei~O Ramos.
\newblock Runaway potentials in warm inflation satisfying the swampland
  conjectures.
\newblock {\em Physical Review D}, 102(10):103522, 2020.

\bibitem{das2020swampland}
Suratna Das, Gaurav Goswami, and Chethan Krishnan.
\newblock Swampland, axions, and minimal warm inflation.
\newblock {\em Physical Review D}, 101(10):103529, 2020.

\bibitem{kamali2020warm}
Vahid Kamali, Meysam Motaharfar, and Rudnei~O Ramos.
\newblock Warm brane inflation with an exponential potential: a consistent
  realization away from the swampland.
\newblock {\em Physical Review D}, 101(2):023535, 2020.

\bibitem{bedroya2020trans}
Alek Bedroya, Robert Brandenberger, Marilena Loverde, and Cumrun Vafa.
\newblock Trans-planckian censorship and inflationary cosmology.
\newblock {\em Physical Review D}, 101(10):103502, 2020.

\bibitem{brahma2020trans}
Suddhasattwa Brahma.
\newblock Trans-planckian censorship, inflation, and excited initial states for
  perturbations.
\newblock {\em Physical Review D}, 101(2):023526, 2020.

\bibitem{brandenberger2020strengthening}
Robert Brandenberger and Edward Wilson-Ewing.
\newblock Strengthening the tcc bound on inflationary cosmology.
\newblock {\em Journal of Cosmology and Astroparticle Physics}, 2020(03):047,
  2020.

\bibitem{dhuria2019trans}
Mansi Dhuria and Gaurav Goswami.
\newblock Trans-planckian censorship conjecture and nonthermal
  post-inflationary history.
\newblock {\em Physical Review D}, 100(12):123518, 2019.

\bibitem{kamali2020relaxing}
Vahid Kamali and Robert Brandenberger.
\newblock Relaxing the tcc bound on inflationary cosmology?
\newblock {\em European Physical Journal C}, 80(4):1--6, 2020.

\bibitem{bernardo2020trans}
Heliudson Bernardo.
\newblock Trans-planckian censorship conjecture in holographic cosmology.
\newblock {\em Physical Review D}, 101(6):066002, 2020.

\bibitem{mizuno2020universal}
Shuntaro Mizuno, Shinji Mukohyama, Shi Pi, Yun-Long Zhang, et~al.
\newblock Universal upper bound on the inflationary energy scale from the
  trans-planckian censorship conjecture.
\newblock {\em Physical Review D}, 102(2):021301, 2020.

\bibitem{brahma2020trans1}
Suddhasattwa Brahma.
\newblock Trans-planckian censorship conjecture from the swampland distance
  conjecture.
\newblock {\em Physical Review D}, 101(4):046013, 2020.

\bibitem{berera2019trans}
Arjun Berera and Jaime~R Calder{\'o}n.
\newblock Trans-planckian censorship and other swampland bothers addressed in
  warm inflation.
\newblock {\em Physical Review D}, 100(12):123530, 2019.

\bibitem{adhikari2020power}
Rathin Adhikari, Mayukh~Raj Gangopadhyay, et~al.
\newblock Power law plateau inflation potential in the rs $ ii $ braneworld
  evading swampland conjecture.
\newblock {\em arXiv preprint arXiv:2002.07061}, 2020.

\bibitem{bravo2020tip}
Rafael Bravo, Gonzalo~A Palma, and M~Sim{\'o}n Riquelme.
\newblock A tip for landscape riders: multi-field inflation can fulfill the
  swampland distance conjecture.
\newblock {\em Journal of Cosmology and Astroparticle Physics}, 2020(02):004,
  2020.

\bibitem{aragam2020multi}
Vikas Aragam, Sonia Paban, and Robert Rosati.
\newblock Multi-field inflation in high-slope potentials.
\newblock {\em Journal of Cosmology and Astroparticle Physics}, 2020(04):022,
  2020.

\bibitem{matsui2019eternal}
Hiroki Matsui and Fuminobu Takahashi.
\newblock Eternal inflation and swampland conjectures.
\newblock {\em Physical Review D}, 99(2):023533, 2019.

\bibitem{kinney2019eternal}
William~H Kinney.
\newblock Eternal inflation and the refined swampland conjecture.
\newblock {\em Physical review letters}, 122(8):081302, 2019.

\bibitem{brahma2019stochastic}
Suddhasattwa Brahma and Sarah Shandera.
\newblock Stochastic eternal inflation is in the swampland.
\newblock {\em arXiv preprint arXiv:1904.10979}, 2019.

\bibitem{wang2019eternal}
Ziwei Wang, Robert Brandenberger, and Lavinia Heisenberg.
\newblock Eternal inflation, entropy bounds and the swampland.
\newblock {\em arXiv preprint arXiv:1907.08943}, 2019.

\bibitem{lin2020topological}
Chia-Min Lin.
\newblock Topological eternal hilltop inflation and the swampland criteria.
\newblock {\em Journal of Cosmology and Astroparticle Physics}, 2020(06):015,
  2020.

\bibitem{randall1999large}
Lisa Randall and Raman Sundrum.
\newblock Large mass hierarchy from a small extra dimension.
\newblock {\em Physical review letters}, 83(17):3370, 1999.

\bibitem{randall1999alternative}
Lisa Randall and Raman Sundrum.
\newblock An alternative to compactification.
\newblock {\em Physical Review Letters}, 83(23):4690, 1999.

\bibitem{gogberashvili2002hierarchy}
Merab Gogberashvili.
\newblock Hierarchy problem in the shell-universe model.
\newblock {\em International Journal of Modern Physics D}, 11(10):1635--1638,
  2002.

\bibitem{kamali2016tachyon}
Vahid Kamali and Mohammad~Reza Setare.
\newblock Tachyon warm intermediate and logamediate inflation in the brane
  world model in the light of planck data.
\newblock {\em Advances in High Energy Physics}, 2016, 2016.

\bibitem{ida2000brane}
Daisuke Ida.
\newblock Brane-world cosmology.
\newblock {\em Journal of High Energy Physics}, 2000(09):014, 2000.

\bibitem{mukohyama2000brane}
Shinji Mukohyama.
\newblock Brane-world solutions, standard cosmology, and dark radiation.
\newblock {\em Physics Letters B}, 473(3-4):241--245, 2000.

\bibitem{sahni2002new}
Varun Sahni and Yuri Shtanov.
\newblock New vistas in braneworld cosmology.
\newblock {\em International Journal of Modern Physics D}, 11(10):1515--1521,
  2002.

\bibitem{brax2004brane}
Philippe Brax, Carsten van~de Bruck, and Anne-Christine Davis.
\newblock Brane world cosmology.
\newblock {\em Reports on Progress in Physics}, 67(12):2183, 2004.

\bibitem{hwang2002cosmological}
Jai-chan Hwang and Hyerim Noh.
\newblock Cosmological perturbations in a generalized gravity including
  tachyonic condensation.
\newblock {\em Physical Review D}, 66(8):084009, 2002.

\bibitem{dimopoulos2018steep}
Konstantinos Dimopoulos.
\newblock Steep eternal inflation and the swampland.
\newblock {\em Physical Review D}, 98(12):123516, 2018.

\bibitem{gibbons1977cosmological}
Gary~W Gibbons and Stephen~W Hawking.
\newblock Cosmological event horizons, thermodynamics, and particle creation.
\newblock {\em Physical Review D}, 15(10):2738, 1977.

\bibitem{gibbons1977action}
Gary~W Gibbons and Stephen~W Hawking.
\newblock Action integrals and partition functions in quantum gravity.
\newblock {\em Physical Review D}, 15(10):2752, 1977.

\bibitem{bekenstein1973black}
Jacob~D Bekenstein.
\newblock Black holes and entropy.
\newblock {\em Physical Review D}, 7(8):2333, 1973.

\bibitem{bousso2006eternal}
Raphael Bousso, Ben Freivogel, and I-Sheng Yang.
\newblock Eternal inflation: The inside story.
\newblock {\em Physical Review D}, 74(10):103516, 2006.

\bibitem{kaul2000logarithmic}
Romesh~K Kaul and Parthasarathi Majumdar.
\newblock Logarithmic correction to the bekenstein-hawking entropy.
\newblock {\em Physical Review Letters}, 84(23):5255, 2000.

\bibitem{ghosh2005improved}
Amit Ghosh and P~Mitra.
\newblock An improved estimate of black hole entropy in the quantum geometry
  approach.
\newblock {\em Physics Letters B}, 616(1-2):114--117, 2005.

\bibitem{medved2004conceptual}
AJM Medved and Elias~C Vagenas.
\newblock When conceptual worlds collide: the generalized uncertainty principle
  and the bekenstein-hawking entropy.
\newblock {\em Physical Review D}, 70(12):124021, 2004.

\bibitem{amelino2004severe}
Giovanni Amelino-Camelia, Michele Arzano, and Andrea Procaccini.
\newblock Severe constraints on the loop-quantum-gravity energy-momentum
  dispersion relation from the black-hole area-entropy law.
\newblock {\em Physical Review D}, 70(10):107501, 2004.

\bibitem{amelino2006black}
Giovanni Amelino-Camelia, Michele Arzano, Yi~Ling, and Gianluca Mandanici.
\newblock Black-hole thermodynamics with modified dispersion relations and
  generalized uncertainty principles.
\newblock {\em Classical and Quantum Gravity}, 23(7):2585, 2006.

\bibitem{meissner2004black}
Krzysztof~A Meissner.
\newblock Black-hole entropy in loop quantum gravity.
\newblock {\em Classical and Quantum Gravity}, 21(22):5245, 2004.

\bibitem{adler2001generalized}
Ronald~J Adler, Pisin Chen, and David~I Santiago.
\newblock The generalized uncertainty principle and black hole remnants.
\newblock {\em General Relativity and Gravitation}, 33(12):2101--2108, 2001.

\bibitem{ali2009discreteness}
Ahmed~Farag Ali, Saurya Das, and Elias~C Vagenas.
\newblock Discreteness of space from the generalized uncertainty principle.
\newblock {\em Physics Letters B}, 678(5):497--499, 2009.

\bibitem{parikh2000hawking}
Maulik~K Parikh and Frank Wilczek.
\newblock Hawking radiation as tunneling.
\newblock {\em Physical Review Letters}, 85(24):5042, 2000.

\bibitem{majumder2011black}
Barun Majumder.
\newblock Black hole entropy and the modified uncertainty principle: a
  heuristic analysis.
\newblock {\em Physics Letters B}, 703(4):402--405, 2011.

\bibitem{majumder2013black}
Barun Majumder.
\newblock Black hole entropy with minimal length in tunneling formalism.
\newblock {\em General Relativity and Gravitation}, 45(11):2403--2414, 2013.

\bibitem{das2008universality}
Saurya Das and Elias~C Vagenas.
\newblock Universality of quantum gravity corrections.
\newblock {\em Physical review letters}, 101(22):221301, 2008.

\bibitem{mead1964possible}
C~Alden Mead.
\newblock Possible connection between gravitation and fundamental length.
\newblock {\em Physical Review}, 135(3B):B849, 1964.

\bibitem{amati1987superstring}
Daniele Amati, M~Ciafaloni, and G~Veneziano.
\newblock Superstring collisions at planckian energies.
\newblock {\em Physics Letters B}, 197(1-2):81--88, 1987.

\bibitem{amati1989can}
Daniele Amati, Marcello Ciafaloni, and Gabriele Veneziano.
\newblock Can spacetime be probed below the string size?
\newblock {\em Physics Letters B}, 216(1-2):41--47, 1989.

\bibitem{maggiore1994quantum}
Michele Maggiore.
\newblock Quantum groups, gravity, and the generalized uncertainty principle.
\newblock {\em Physical Review D}, 49(10):5182, 1994.

\bibitem{Scardigli:2016pjs}
Fabio Scardigli, Gaetano Lambiase, and Elias Vagenas.
\newblock {GUP parameter from quantum corrections to the Newtonian potential}.
\newblock {\em Phys. Lett. B}, 767:242--246, 2017.

\bibitem{Kanazawa:2019llj}
T.~Kanazawa, G.~Lambiase, G.~Vilasi, and A.~Yoshioka.
\newblock {Noncommutative Schwarzschild geometry and generalized uncertainty
  principle}.
\newblock {\em Eur. Phys. J. C}, 79(2):95, 2019.

\bibitem{Buoninfante:2019fwr}
Luca Buoninfante, Giuseppe~Gaetano Luciano, and Luciano Petruzziello.
\newblock {Generalized Uncertainty Principle and Corpuscular Gravity}.
\newblock {\em Eur. Phys. J. C}, 79(8):663, 2019.

\bibitem{pedram2012higher}
Pouria Pedram.
\newblock A higher order gup with minimal length uncertainty and maximal
  momentum.
\newblock {\em Physics Letters B}, 714(2-5):317--323, 2012.

\bibitem{pedram2012higher2}
Pouria Pedram.
\newblock A higher order gup with minimal length uncertainty and maximal
  momentum ii: Applications.
\newblock {\em Physics Letters B}, 718(2):638--645, 2012.

\bibitem{bandyopadhyay2018thermodynamic}
Tanwi Bandyopadhyay.
\newblock Thermodynamic prescription of cosmological constant in the
  randall-sundrum ii brane.
\newblock {\em Advances in High Energy Physics}, 2018, 2018.

\end{thebibliography}
\bibliographystyle{unsrt}

\end{document}